# Toward Timed-Release Encryption in Web3 – An Efficient Dual-Purpose Proof-of-Work Consensus

Fanghao Yang, *Member, IEEE*, Xingqiu Yuan

**Abstract** Many existing timed-release encryption schemes uses time-lock puzzles to avoid relying on a trusted timeserver or a key holder which could be a weak spot in data security. However, it is unavoidable to consume massive computing power for solving time-lock puzzles and it is difficult for encryptors to predict the amount of time to solve a puzzle by decryptors. In this study, an efficient dual-purpose proof-of-work consensus allows users to release a time-locked content, which is encrypted by an asymmetric key encryption scheme on a blockchain, without trust in any third-party agents. The release time is predictable as the block time in a proof-of-work blockchain is adaptively controlled. The mining work is repurposed so that once a new block was mined on the blockchain network, time-lock puzzles were also solved immediately. No additional work is required to reveal the time-locked contents and the encryption is secured by monetary incentive mechanisms since it would be very costly to arrange an attack attempt, which must overtake the total hash rate of the whole blockchain network.

## I. INTRODUCTION[1]

It is highly expected that new digital economy would boom inside the metaverse beyond the limitation of the physical world. It would be important to design efficient data exchange protocol for protection of privacy in the metaverse, where everything is digitized as a piece of data. Thus, it is believed that the foundation of the metaverse shall be built on top of decentralized networks (Web3) and decentralized autonomous organizations (DAOs) shall be the major body of the governance within the metaverse [1]. Many critical social economic behaviors within the metaverse, e. g. voting in DAOs and blind bidding in the Web3 market, requires delayed release of confidential data or actions. To achieve those applications, a decentralized timed-release encryption scheme needs to be developed with blockchains on Web3. Although the timed-release mechanism can be implemented by a trusted agent as it was seen in Web 2.0, the confidential contents cannot be securely hidden from the third-party agents so that the hidden contents may be leaked by whoever holds the data or the key. To address this issue, many researchers proposed the use of multiple trusted agents [2], [3] instead of a single agent to enhance the data securities and avoid early data leakage, however, it requires to build many secured messaging channels between multiple agents which increase the complexity of the whole system and computation & communication overheads. In all, the methods with trusted agents cannot be well adopted and a third-party agent cannot be trusted, if the encrypted message contains financially or socially valuable data. Instead, the time-lock puzzle requires time-consuming computations to solve, and the hidden data in the time-lock puzzle can only be revealed after certain amount of time. There is no need to trust any third-party to reveal the hidden content over elapsed time. However, to solve those puzzles and decrypt the message, a significant amount of computation, which implies a high cost of operation, is required. Moreover, it is difficult to accurately predict the elapsed time for finding the puzzle solution because it depends on the speed of the computer hardware which participants in the problem solving and the computing power could be augmented or the computer algorithm/architecture could be optimized over the time after the release of the time-lock puzzle (e. g. more processors or a new type of processor may be utilized to solve the problem and release the message earlier than the prediction).

In recent years, a proof-of-work (PoW) blockchain, for example, Bitcoin, was adopted as a decentralized ledger to store the token transaction history. It is worth mentioning that the block time, which is the time required to create a new block on the blockchain, is a controlled to be stable since there is an adaptive control mechanism within the mining algorithm and consensus, which adjust its mining difficulty based on the average block time of the current time window. Since the deployment and builds of new mining equipment are based on the physical production rate of computer hardware which is relatively stable within a short time window (for example, a day or a week), if the whole mining network grows to a very large scale, the mining hash rate, which represents the computing power in the whole network, is as stable as the computer production/deployment rate. It was observed that the whole Bitcoin network can adapt to gradual hash rate changes over time and maintain stable block time [4]. This stable block time can be the decentralized clock tick to trigger the release of hidden contents with a decentralized timed-release encryption scheme for Web3 applications.

---

[1] Fanghao Yang is with the NexToken Technology, Princeton, NJ 08852 USA (e-mail: yangfanghao@gmail.com).
Xingqiu Yuan is with Argonne National Laboratory, Lemont, IL 60439 USA (e-mail: xyuan@anl.gov).

## II. RELATED WORK

Jia Liu, Tibor Jager and et al, uses above-mentioned block time $\Delta t$ as a reference clock tick to design a timed-release encryption scheme mainly based on the witness encryption [5]. In this scheme, the Bitcoin blockchain, or any other alternative PoW blockchain, works as a witness to encrypt messages as current block height $x$. To decrypt the message, a witness (Bitcoin Blockchain) with future block height $x'$ would be required to compute with a cryptographic multi-linear map. The encrypted message is automatically released to public once the Bitcoin blockchain grow to certain block height $x'$. Due to that the block time $\Delta t$ is referred as a stable clock tick, the whole blockchain network can be used as a computational reference clock to determine the release time of the hidden message so that the time elapse for releasing the message would be $\Delta t(x' - x)$. However, the construction of this theoretical scheme is not only complex but also impractical. Without using a zero-knowledge proof system SNARKs [6], the verification of mining results is designed to be inefficient [5]. Even with the SNARKs, yet a huge and complex formula would be too costly to verify the mining results. Consequently, there is no implementation for above-mentioned encryption scheme to achieve a practical timed-release mechanism.

Using the stable block time $\Delta t$ as a reference clock tick, a decentralized and predictable timed-release mechanism was also proposed by Sang-Wuk Chae and et al without using any complex encryption scheme [7]. In their study, discrete log problem (DLP) was utilized to implement timed-release encryptions. They suggested a chained block structure to store encrypted messages within blocks and each block contains the corresponding public key for message encryption. The public key in a DLP-based crypto system were generated in a deterministically pseudo-random manner, so that a list of public keys is pre-determined before the corresponding blocks were mined. The mining algorithm, which is a SHA-256 hash function, is essentially equivalent to a generator of pseudo-random numbers, therefore, the generated hash results are also reduced/mapped to an integer and tested as the private key against the pre-determined public keys of the above crypto system. Once the private key was found, the block was sealed on the blockchain as part of PoW consensus, and the encrypted message contained in current new block is then decrypted publicly. The mining difficulty can be dynamically adjusted by choosing a prime number $p$ with different bit length based on average block time of recent blocks. However, their mining approach is analogous to a brute-force attack on DLP encryption, which has time complexity $O(2^n)$ where $n$ is the bit length of the private key in DLP crypto system. For finding the corresponding private key, their approach is significantly less efficient than Pohlig-Hellman or Pollard-Rho algorithms [8] which has time complexity $O(e^{0.5n})$. If using efficient algorithms to search the private key, the attacker may only take a very small fraction of the total computing power of the whole blockchain network to get early access to the encrypted message. Their approach cannot secure valuable messages and prevent data leaks from low-cost attacks so that it is not safe to deploy in real-world applications.

The goal of this study is to enhance the security of above-mentioned encryption scheme toward a practical

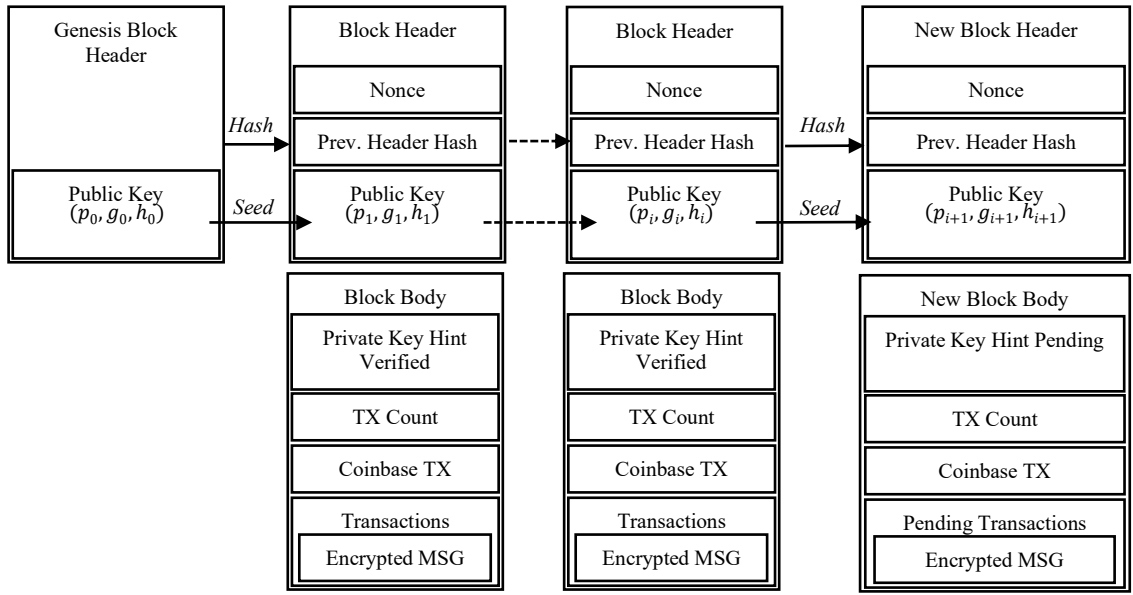

Fig. 1 Diagram of the time-release blockchain. *Hash* represents the hash function to calculate the hash value of previous block header. *Seed* represents the random number generator uses the sum of previous public key as the seed

decentralized timed-release blockchain. It is necessary to find a new mining algorithm which not only meets the requirements of PoW consensus but also efficiently solve the time-lock puzzle so that the monetary incentive mechanism would prevent attackers from using significantly more efficient algorithms to find the private key and decrypting the valuable message ahead of time.

### III. BLOCKCHAIN WITH TIMED-RELEASE ENCRYPTION

In this study, the chained block structure is similar to the research of Sang-Wuk Chae and et al [7]. Besides all basic fields in a PoW blockchain (e. g. timestamp, Merkle tree root), the block contains three pieces of data to ensure the immutability and PoW consensus as shown in Fig. 1.

#### A. Hash of previous block header

like Bitcoin or any other PoW blockchains, the hash result of previous block header makes all data and history on blockchains immutable, any changes on previous block must regenerate the nonce or redo the work.

#### B. Public key

Each block header contains the corresponding public key to encrypt the hidden message to be released in current block time. The public key and private key are based on DLP crypto system. For example, if a user intends to release a piece of message in a future block time, the message may be encrypted by the corresponding public key. If the publisher of the hidden data wants to make the data immutable and publicly available after the release time (e. g. DAO voting), the message may be written into the blockchain as part of immutable transaction metadata as shown in Fig. 1.

As shown in Fig. 1, the public key for each block is chained by using previous public key as the seed value for the random number generator. A deterministic random number generator (e. g. Mersenne twister pseudorandom number generator [9]) may be used to generate a list of chained public key for a blockchain. For example, assume that the public key of the corresponding block is known, during the generation of the subsequent public key, the sum of previous public key $(p_i, g_i, h_i)$, where $p_i$ is a safe prime number, $g_i$ is the generator and $h_i$ is the element in the subgroup of the cyclic group of the generator $g_i$, is used as a seed for the generation of a new safe prime number $(p_{i+1}, g_{i+1}, h_{i+1})$. The first public key is hard coded in the genesis block; therefore, anyone can generate the same list of chained public keys in any time without knowing the private key.

#### C. Nonce

Nonce value serves dual purposes: like Bitcoin, it is a solution for a known hard problem (may not be proved as NP-hard). Unlike Bitcoin, the private key (not necessary the nonce value itself) shall be easily derived from the nonce value in constant time. Once the nonce value is found, the private key is also derived for releasing message within current block time.

### IV. TIMED-RELEASE MECHANISM

As descripted in the previous section, the user, who is going to encrypt a message and schedule the release of the encrypted message on the blockchain, shall determine the block height of its release time using the equation below, where x is current block height, $t'$ is the release time and t is current time.

$$x' = \frac{(t' - t)}{\Delta t} + x \qquad (1)$$

Then, the block height of the release time $x'$ is derived and its corresponding public key is generated using the method in the above section. With this public key, the message is then encrypted and published into the blockchain as an attached metadata of a transaction, which pays a fee to the miner who is supporting the security of the blockchain.

When the block at the height $x'$ is mined on the release time $t'$, the corresponding private key is then automatically found by below mining mechanism so that the message is decrypted and released publicly. Anyone can unlock and acquire the released message by accessing the immutable blockchain data.

### V. MINING MECHANISM AND ALGORITHM

The cornerstone of above mentioned blockchain system is the mining algorithm which enable dual purposes of the mining work. The dual purposes mean that the mining work not only form the network consensus which ensure the immutability of the encrypted message (and token transaction history) but also practically implemented the time release mechanism by solving the time-lock puzzle in the release time.

In Bitcoin, the SHA256 algorithm, which is represented as a function SHA256(), essentially serves as a part of a pseudo-number generator looping through all possible nonce values embedded in the block header to solve the mining challenge, which is searching for a few consecutive zeros in its hash results. As a proof of work, this mining target, nonce, has no better way to find than brutal force computing. In this study, instead of looking for the same nonce as Bitcoin uses in its mining work, the blockchain protocol will verify whether the hash results of a randomly generated nonce could derive the private key for the corresponding public key of current reference block time. It is assumed that the public key for the candidate block is known as $(p, g, h)$, if using a safe prime number $p$, the order $n = (p - 1)/2$ is also a prime number, and we have the header hash function of the candidate block, where *int*() represent a cast function convert the output of a SHA256 function into a large integer and *header*(*nonce*) represent a block header contains a variable nonce.

$$\text{hash}(nonce) \equiv \\ int\left(\text{SHA256}\left(\text{SHA256}(header(nonce))\right)\right) \qquad (2) \\ (\bmod p)$$

The mining algorithm is based on pollard rho method[10]. Thus, we need to construct a set of mapping functions for pseudo random walk in a cyclic graph. The mapping functions should always ensure $y_i = g^{a_i} h^{b_i}$ where $a_i$ and $b_i$ are intermediate parameters for finding the private key with birthday attacks. The nonce in the block header also casts to an integer number $y_i$. To simplify this early study, Floyd's cycle-finding algorithm is used to find the collision, but other cycle-detection methods may be also used for better space & time efficiency. Thus, two sets of $(y_i, a_i, b_i)$ from the same cyclic group $G$ walks in the mapped cyclic graph with different paces until they were collided.

It is worth noting that the nonce value $y_i$ is also contained within the block header as shown in Fig. 1. Assuming the other parts in the header were determined while mining, the only variable in the hash function is the nonce value, thus, the equation 2 is equivalent to a function of $y_i$ noted as $\text{hash}(y_i)$ for a concise statement. Here, we have three mapping functions to generate a set of intermediate parameters from the hash results. The mapping uses three partitions where $G = S_0 \cup S_1 \cup S_2$.

The first mapping function for $a$:
$$a_{i+1} = g(a_i, y_i)$$
$$\equiv \begin{cases} a_i \cdot \text{hash}(y_i) (\text{mod } (p-1)), & \text{hash}(y_i) \in S_0 \\ a_i + \text{hash}(y_i) (\text{mod } (p-1)), & \text{hash}(y_i) \in S_1 \\ a_i (\text{mod } p), & \text{hash}(y_i) \in S_2 \end{cases} \quad (3)$$

The second mapping function for $b$:
$$b_{i+1} = h(b_i, y_i)$$
$$\equiv \begin{cases} b_i \cdot \text{hash}(y_i) (\text{mod } (p-1)), & \text{hash}(y_i) \in S_0 \\ b_i (\text{mod } p), & \text{hash}(y_i) \in S_1 \\ b_i + \text{hash}(y_i) (\text{mod } (p-1)), & \text{hash}(y_i) \in S_2 \end{cases} \quad (4)$$

In above equations 3 and 4, $(p-1)$ is the results of Euler's totient function $\varphi(p)$, which ensure the node $(y_i, a_i, b_i)$ is always in the same group $G$.

The third mapping function for $y$:
$$y_{i+1} = f(y_i)$$
$$\equiv \begin{cases} y_i^{\text{hash}(y_i)} (\text{mod } p), & \text{hash}(y_i) \in S_0 \\ g^{\text{hash}(y_i)} \cdot y_i (\text{mod } p), & \text{hash}(y_i) \in S_1 \\ h^{\text{hash}(y_i)} \cdot y_i (\text{mod } p), & \text{hash}(y_i) \in S_2 \end{cases} \quad (5)$$

The mining algorithm is to find the cycle in the mapped graph from the group $G$. A walking step means that mapping functions were used once to calculate the next node $(y_{i+1}, a_{i+1}, b_{i+1})$ from current node $(y_i, a_i, b_i)$ in the cyclic graph with step length of a single edge.

| Algorithm 1: Mining Algorithm |
| --- |
| **Input**: public key $(p, g, h)$ |
| **Output**: a tuple of integer $(a^1, b^1, a^2, b^2)$ and nonce value y |
| 1    $n \leftarrow (p-1)/2$ |
| 2    Prepare initial value for $a_0$, $b_0$ and $y_0$:     $a_0^{1,2}, b_0^{1,2} \leftarrow$ random integers between 0 and n     $y_0^{1,2} \leftarrow g^{a_0} h^{b_0}$ |
| 3    $i \leftarrow 1$ |
| 4    **while** $i < n$ **do** |
| 5        walk single step from $(y_{i-1}^1, a_{i-1}^1, b_{i-1}^1)$ to $(y_i^1, a_i^1, b_i^1)$ using equation 3 - 4 |
| 6        walk double steps from $(y_{i-1}^2, a_{i-1}^2, b_{i-1}^2)$ to $(y_i^2, a_i^2, b_i^2)$ using equation 3 - 4 |
| 7        **if** $y_i^1 = y_i^2$ **do** |
| 8            derive the private key from $(a_i^1, b_i^1, a_i^2, b_i^2, n)$ as the pollard rho method [10] |
| 9            **if** private key pairs with the public key **do** |
| 10               **return** $(a_i^1, b_i^1, a_i^2, b_i^2, y_{i-1}^2)$ |

Knowing the solution tuple $(a^1, b^1, a^2, b^2)$ and the public key, the private key can be easily derived using classical textbook method [10]. The most recent nonce value $y_{i-1}$ is also returned to be contained within the block header. The nonce value $y_{i-1}$ is unique since it maps to the intermediate parameter $y_i$ with the mapping equation 5 noted as $f(y_{i-1})$ and it shall meet the condition $y_i \equiv g^{a_i} h^{b_i} (\text{mod } p)$. The time complexity of this mining algorithm is $O(\sqrt{p})$ like Pollard Rho method. The mining difficulty may be adjusted by using a safe prime number with different big length. If a mining node found the solution and the nonce value, it can broadcast the solution tuple $(a^1, b^1, a^2, b^2)$ and the candidate block to the whole network for validation.

Then, we have the following Algorithm 2 to validate the newly sealed block. The mapping function (equation 4) is used to verify the nonce value. Since there is no loop or complex formula in this validation algorithm, the time complexity is $O(1)$, which is a significant improvement over existing study on witness encryption [5].

| Algorithm 2: Mining Validation Algorithm |
| --- |
| 1    for all nodes in network do |
| 2        **if** receive a new candidate block and its solution tuple of integer $(a^1, b^1, a^2, b^2)$ **do** |
| 3            $y_{i-1} \leftarrow$ nonce value in the block header |
| 4            **if** $f(y_{i-1}) \equiv g^{a^1} h^{b^1} (\text{mod } p)$ and $f(y_{i-1}) \equiv g^{a^2} h^{b^2} (\text{mod } p)$ **do** |
| 5               derive the private key from $(a^1, b^1, a^2, b^2, n)$ as the pollard rho method [10] |
| 6               **if** the private key pairs with the public key **do** |
| 7                   add the candidate block to the blockchain |

## VI. ANALYSIS OF DATA SECURITY

In this section, two types of attacks are analyzed against above-mentioned algorithms and methods. The following analysis is based on a hypothesis that it is not significantly less difficult to solve the equation 5 reversely than to solve the puzzle itself. In another word, there is no easy solution (time complexity is significantly less than $O(\sqrt{p})$) to find $y_{i-1}$ if knowing $y_i$. This hypothesis seems valid because equation 5 combines discrete log function and SHA-256 hash function. Both functions are known to be one-way functions [11], [12].

The first type of attacks is attack on immutability of the ledger data while hackers would try to modify data on the blockchain including the encrypted message and transaction history. Once the mining work is finished, the solution tuple is broadcast into the public network, since then, attackers shall also know the existing solution tuple of the time-lock puzzle. If the block is modified by the attacker, the hash function hash($y_{i-1}$) will generate a different hash result so that the result of $f(y_{i-1})$ in equation 5 would be also different (noted that $f(y_{i-1})$ depends on hash($y_{i-1}$) ). Therefore, the attacker must either find a new nonce $y_{i-1}$ or modify other parts of the block (e. g. timestamp) to replicate the same hash result to pass the step 4 in the validation algorithm (Algorithm 2) so that the hacked ledger data would be accepted to the blockchain. Since $y_{i-1}$ and hash($y_{i-1}$) could be any integer number between 0 and $p$, with brutal force attacking, the time complexity is $O(p)$ to meet the condition in step 4 of validation algorithm.

The attacker may also solve the puzzle again by regenerating the solution tuple with a set of new $(a^1, b^1, a^2, b^2)$ with a new nonce value $y_{i-1}$. Any changes in the block data, which is modified by the attacker, will induce the change of the hash function hash($y_{i-1}$) (equation 2), consequently, generate a new set of mapping functions (equation 3 - 5). The attacker, who is trying to modify the block data, must generate a new cyclic graph and find a new collision by spending the same amount work as mining a new block with time complexity of $O(\sqrt{p})$, approximately.

In general, the purposed mining algorithm ensures the PoW consensus in the blockchain so that the mining mechanism works as it should be.

The second type of attacks is to release the encrypted message before its corresponding block is mined. The premature release of the message will damage the time-release mechanism and suppress the usage of this blockchain system. To attack the DLP-based encryption, the lower limit of the time complexity is $O(\sqrt{p})$ [13] which ensure that a significant amount work must be done to prematurely release the message. Thus, the cost of attacking on an encrypt message is no less than mining a new block. Due to the high attack cost, the encrypted message is safe to be distributed publicly on the blockchain. This is a major improvement over early study [7] whose encryption message can be easily attacked with a small fraction of mining work.

## VII. Token Economy

Token economy is briefly discussed here to demonstrate the incentive mechanism of mining since no miner will do work for free. Like Bitcoin, the miner is driven by a token economy system. The user of the blockchain pays the miner with a fee for timely releasing their encrypted message. Moreover, the miner who found the nonce to seal the new block (and the private key for corresponding public key) is rewarded with several newly "mined" tokens as receiving the coinbase transaction.

The tokens may be exchanged from miners to end users as a utility tokens and users will pay for the use of the above mentioned time-release blockchain network. Users who are paying higher fees (more tokens) per size of their contents will get their encrypted message written into the blockchain in time with higher priority since each block has limited size (e. g. 4 MB). The time release mechanism may also be used by off-chain data; however, the immutability of off-chain data and its operation history is not secured by PoW consensus. In many applications, e. g. a decentralized voting system on Web3, eligibility of the voting data is determined by the time of its creation and is not allowed to be modified once it was created. Thus, the users who need their data to be immutable and later published, shall put their data on a public blockchain and pays a fee with utility tokens.

Comparing to existing approaches using time-lock puzzles for each encryption, the cost of unlocking a single piece of content is relatively low and affordable (in the same scale of the transaction fee in the Bitcoin network). Because the release of encrypted data is in a batch after a new block is mined, the cost for a single use is a fraction, which is the cost of mining divided by the number of total users, while the timed-release encryption is secured by the total hash power of the whole network.

The mining efficiency may be improved with a newly purposed algorithm and the total hash power in this network could be increased from new miners because of monetary incentives from mining. The cost of attacking the timed-release encryption would be so high that the attacker would rather to earn mining rewards safely with the same amount computing power. As a result, the on-chain data is prevented from premature leaks.

## VIII. Conclusion

In this study, a dual-purpose mining algorithm were proposed to enhance the security toward a practical time-release blockchain on Web3. The mining work has been repurposed as decryption work against the encrypted message so that no extra cost would be necessary to release the time-locked message. The timing of release is based on a reference clock tick as the block time in the difficulty-adjustable mining process so that the message can be released stably and accurately.

The message is not only encrypted/released in a timely manner but also immutable if being put on a public blockchain. This property is critical for many real-world applications (e. g. decentralized voting) and may create long-term impact on economics and politics.

This study still has some potential challenges for future works, for example, the difficulty adjustment depends on the bit length of prime number. Since the mining difficulty is hard to predict and adjusted in real-time, the generation of the public key of future blocks are also unpredictable. We need to either find a way to stabilize total mining hash rate so that the mining difficulty would be predictable in a reasonable time frame, or multiple parallel chains of public keys with different bit lengths may be used for redundant encryption of messages so that any of private keys were found, the message could be released.

Moreover, the mapping functions (Equation 3 - 5) in this study may be further simplified to provide the same level of protection so that the mining efficiency get further optimization, provided that the encryption attacking algorithm (e. g. Pollard-rho algorithm) shares the same big O notation with our mining algorithm.

APPENDIX

The algorithms and methods in this study are implemented and demonstrated in Python as a minimum viable demo. The demo software may be accessed with a public GitHub repo (https://github.com/yangfh2004/Time-Release-Blockchain)

**Fanghao Yang** (M'16) was born in Hunan, China. He received the B.Eng. degree in measurement and instrumentation from Beihang University, Beijing, China, in 2008, the Ph.D. degree in mechanical engineering from the University of South Carolina, Columbia, SC, USA, in 2013, and the M.S. degree in computer science from the Georgia Institute of Technology, Atlanta, GA, USA, in 2019. From 2014 to 2016, he was a Post-Doctoral Researcher with IBM Research, Yorktown Heights, NY, USA. He was working on system modeling and measurement & control of data center cooling system. From 2016 to 2020, he was a Scientific Application Software Engineer with the Princeton Plasma Physics Laboratory, Princeton, NJ, USA. He was working on experimental data analysis and management, high-performance computing for tokamak simulation, and real-time system of plasma control. Since 2020, he was working on development of decentralized system and its applications in Web3.